# Numerical Modeling of Ion Transport and Adsorption in Porous Media: A Pore-scale Study for Capacitive Deionization Desalination

Min Liu[1], John Waugh[2,3], Siddharth Komini Babu[2], Jacob S. Spendelow[2], Qinjun Kang[1*]

1 Earth and Environmental Science Division, Los Alamos National Laboratory, Los Alamos, NM 87545, USA

2 Materials Physics and Applications Division, Los Alamos National Laboratory, Los Alamos, NM 87545, USA

3 Material Science Department, Vanderbilt University, Nashville, TN 37235

\* Corresponding author

## Abstract

A pore-scale model is presented to simulate the dynamic ion transport and adsorption processes in porous electrodes used for capacitive deionization (CDI). The Stokes equation governing water flow is solved using the lattice Boltzmann method and Nernst-Planck equation describing ion transport is solved using the finite volume method. The ion adsorption process is considered at the surface of carbon electrodes. After validation against analytical solutions and published results, the model is used to study the coupled water flow, ion transport and adsorption in both two-dimensional and three-dimensional porous CDI electrodes at the pore scale. The effect of electrode microstructure, electrical potential and flow velocity on the adsorption processes is quantitatively investigated, and the relative importance of various parameters is determined. The presented model provides a numerical tool to quantitatively analyze the ion transport and adsorption in porous electrodes. It can help improve the fundamental understanding of the adsorption processes in porous electrodes for capacitive deionization.

# 1. Introduction

The increasing need for clean and drinkable water, in the face of a changing climate and growing global population, has become a critical challenge. Limited freshwater resources are insufficient to meet this growing demand [1–3]. Due to the abundance of seawater and brackish water resources, desalination becomes an attractive approach of great potential to produce clean and potable water.

Desalination techniques of removing inorganic ions mainly include thermal (distillation), pressure (reverse osmosis) and electrical driven (capacitive deionization and electrodialysis) methods. In distillation, the freshwater is obtained by firstly vaporizing the seawater or brackish water and then condensing the steam. The application of this technique is limited due to high energy consumption. In contrast, reverse osmosis (RO) is a pressure-driven and membrane-based method which employs semipermeable membranes to selectively remove the salt from seawater. However, the requirement of high pressure increases the operational cost while the transport through membrane limits the production efficiency of potable water [4]. Unlike distillation and RO, the capacitive deionization (CDI) method enables removal of the minor constituent (salt) from the major constituent (water) based on electrochemical adsorption in the active carbon electrodes, and is therefore particularly attractive for purification of dilute saltwater resources (i.e., brackish water) [5]. During water desalination using CDI, seawater flows through/by the porous electrodes of the CDI cell. Salt ions such as $Na^+$ and $Cl^-$ are driven by the electrostatic force and adsorbed at the surface of the porous active carbon materials in the electrodes and thus removed from the water [1,6].

In the past ten years, there have been an increasing number of studies on CDI to improve performance of desalination [1,3,7–9]. Different CDI cell architectures have been developed, aiming to increase the ion adsorption capability and reduce energy requirement. Among these, two of the most popular CDI cell architectures are flow-by and flow-through CDI [10]. These two systems both consist of two porous electrodes made of adsorptive carbon materials. In a flow-through cell, as shown in Figure 1, the feed water flow is perpendicular to the porous electrodes. An electrical potential is applied between the porous electrodes, driving charge on the surface of the carbon materials. The ion transport and adsorption in CDI cells occur in the pore space within the electrodes. The movement of ions is affected by advection, diffusion, and electrostatic force.

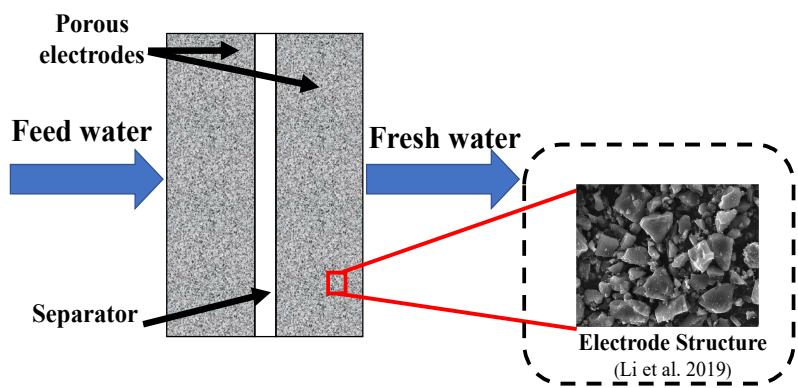

Figure 1 Schematic representation of the architecture of a flow-through CDI cell. Image of electrode structures is reproduced from Li et al., (2019) [11]

Apart from the architectures of the CDI cell, the high ion transport resistance and low ion adsorption capacity are currently the main limitations of CDI performance [12]. Extensive research on improving ion adsorption capacity and decreasing ion transport resistance has been conducted [12–17]. For instance, Porada et al. (2012) [12] constructed CDI cells with different activated carbon materials and compared their salt adsorption capacity. They found that carbon materials with well-defined sub-nanometer pore sizes show significantly higher salt adsorption capacity at the same voltage, indicating the potential of carbide-derived carbon materials for energy-efficient water desalination. Qu et al. (2015) [8] characterized the electric resistance in the CDI system by proposing measurable figures of merit. They measured resistive components in circuit models and found contact pressure between porous electrodes and current collectors can lead to low contact resistance. Nevertheless, these studies were mostly focused on experimental investigations, which are often time consuming and expensive. Alternatively, physics-based modelling can be performed under different operating conditions to find critical parameters more quickly and less expensively than experiments. Thus, the physics-based model that accounts for transport and adsorption of ions in the CDI electrodes has become a promising method to help improve the desalination performance.

Recently, several numerical models have been developed to quantify and analyze ion transport and adsorption in porous CDI electrodes [9,18–22]. Notably, Liu et al. (2019) [22] developed a pore-scale model accounting for ion transport and adsorption in porous electrodes of a CDI cell. They used this model to investigate the transient ion adsorption at the pore scale and studied the effects of porous structures on the ion adsorption process. They found that the inlet velocity does not affect the final ion adsorption amount, but changes the time to reach the final steady state. Their results also showed that porous structures with a moderate particle aggregation degree have the best capacitive deionization performance. Salamat and Hidrovo (2020) [3] studied the impact of micropores on adsorption/desorption of ions in a CDI cell. They included the sorption resistance and non-electrostatic attractive forces into their model and compared the simulation results with experiments using a circular CDI cell operating under various conditions. Their results showed that neglecting the electro-sorption resistance can result in up to 50% overestimation of the energy efficiency and overall desalination performance.

Coupled ion transport and adsorption in porous media have also been extensively studied at the continuum scale, especially in the earth and environmental sciences community. Tournassat and Steefel (2019) [23] and Steefel and Tournassat (2021) [24] investigated the diffusion and adsorption process in charged nanoporous media. They considered a continuum scale/micro-continuum treatment of electrostatic effects on transport and captured the adsorption behaviours at charged mineral surface.

However, very few models have considered explicit pore structures and adsorption-transport behaviours in 3D fibrous media. Besides, no previous numerical study has been performed directly on the real geometries of carbon electrodes obtained from scanning electron microscope (SEM) images. Herein, a pore-scale model based on direct numerical simulations is presented to examine ion transport and adsorption in CDI systems under different conditions. The effect of pore structures and electrical potential on ion adsorption in porous electrodes are quantitatively determined. The results are interpreted using ion concentrations and dimensionless quantities.

## 2. Numerical model setup

In this section, the partial differential equations governing fluid flow, ion transport and adsorption in porous electrodes are first presented. Then the numerical methods to solve these equations are briefly introduced.

Laminar and incompressible fluid flow at low Reynolds number is assumed in the porous electrodes. Thus, the continuity and the Stokes equations are considered [25]

$$\nabla \cdot \boldsymbol{u} = 0, \tag{1}$$

$$\nabla p = \mu \nabla^2 \boldsymbol{u}, \tag{2}$$

where $\boldsymbol{u}$ ($m \cdot s^{-1}$) is fluid velocity vector, p (Pa) is pressure, and µ ($kg \cdot m^{-1} s^{-1}$) is dynamic viscosity of the fluid.

In current study, adsorption of $Na^+$ and $Cl^-$ ions from saline water is considered. The transport of ions is governed by the Nernst-Planck equation [26]

$$\frac{\partial C_i}{\partial t} + (\boldsymbol{u} \cdot \nabla) C_i = \nabla \cdot (D_i \nabla C_i) + \frac{E z_i D_i}{K_B T} \nabla \cdot (C_i \nabla \psi), \tag{3}$$

where $C_i$, $D_i$ and $z_i$ represent the local concentration ($mol \cdot m^{-3}$), diffusion coefficient ($m^2 \cdot s^{-1}$), and ion valence of the *i*-th ion, respectively. $E$, $K_B$, $\psi$, T represent the value of one proton charge (C), Boltzmann constant ($JK^{-1}$), electrical potential (V) and temperature (°C), respectively. The distribution of electrical potential obeys the following Poisson equation:

$$\nabla^2 \psi = -\frac{\rho_e}{\varepsilon_r \varepsilon_0} \tag{4}$$

where $\varepsilon_0$ is the vacuum electric permittivity ($C \cdot V m^{-1}$) and $\varepsilon_r$ is the dimensionless dielectric constant. Dirichlet boundary condition is enforced at the liquid-solid interface. A constant value of potential is applied at the solid surface. Initially, the potential in fluid phase is set to zero. The $\rho_e$ denotes the net charge density ($C \cdot m^{-3}$) and is related to the ion concentration:

$$\rho_e = \sum_i F z_i C_i \tag{5}$$

where $F$ is the Faraday constant. The adsorption on the surface of carbon materials in the porous electrodes is described by

$$D_i \frac{\partial C_i}{\partial \underline{n}} = K_a C_i, \tag{6}$$

with the adsorption constant $K_a$ ($m \cdot s^{-1}$). The linear reaction kinetics has been used in many studies to describe the adsorption process [22,27–29]. In current study, the concentration of the injected NaCl is much lower than the saturation. During the desalination/adsorption process, the adsorption rate is hence determined by the concentration at the solid surface at a given adsorption rate constant.

The governing equations for fluid flow are solved via the lattice Boltzmann method (LBM). We apply the direct numerical simulations methods which is performed directly on the cartesian grids of the geometries of images without meshing process. We use structured marker-and-cell (MAC) gridding such that each grid block in the numerical simulation represents a voxel on the micro-CT image [30]. The size of each grid is determined by the resolution of images.

The LBM solves for fluid flow by applying the discrete Boltzmann equation,

$$f_i(x + e_i, t + 1) - f_i(x, t) = \Omega_i(x, t), \qquad (7)$$

where $f_i$ is the particle distribution function in the $i$-direction, $e_i$ is the microscopic velocity in the $i$-direction and $\Omega$ is the collision operator. We apply the Bhatnager-Gross-Krook (BGK) model utilizing a linear collision operator where the particle distribution function is expanded around its equilibrium value,

$$\Omega_i(x, t) = -\frac{1}{\tau}\left(f_i(x, t) - f_i^{eq}(x, t)\right), \qquad (8)$$

where $f_i^{eq}$ is the local equilibrium state and $\tau$ is the relaxation time parameter. The macroscopic fluid flow quantities, density and velocity, are calculated using $\rho$.

A no-flow boundary condition is applied on solid-fluid interfaces which in the LBM is realized through the bounce-back rule [31]. The pressure gradient acting on the fluid is simulated by a body force and the mirror image boundary conditions are applied on the plane perpendicular to the flow direction.

With an appropriate choice of $e_i$ and $f_i^{eq}$, Eq. (7) can be proved to recover the incompressible Newtonian Navier-Stokes equations through a Chapman-Enskog procedure [32], with fluid density and velocity given by $\rho = \sum_i f_i$ and $u = \frac{1}{\rho}\sum_i e_i f_i$, respectively. For the commonly used D2Q9 (two-dimensional nine-speed) and D3Q19 (three-dimensional nineteen-speed) lattice models, the kinematic viscosity is calculated with the relaxation time as $v = \frac{1}{3}(\tau - \frac{1}{2})$ [31,33].

An implicit finite volume method is employed to discretize the governing equations for ion transport [34–37]. The details of the numerical methods can be found in [36,38].

## 3. Model validation

For the model validation of ion transport and adsorption due to advection and diffusion developed by Liu and Mostaghimi (2017) [38], we compare our simulation results with the previous studies by Zhou et al. (2015) [39]. An identical system is used in both Zhou et al. (2015) [39] and our validation simulations. The model simulates the fluid flow through a channel with ion adsorption at the bottom surface (as shown in Figure 2a). The size ($L \times L$) of the domain is 320×320 in lattice units. A constant concentration 1.0 is enforced at the inlet and surface adsorption obeying the first-order kinetics is considered at the bottom wall of the channel. No-slip boundary condition is applied at the upper wall. The adsorption rate constant is set as 1.0, while the diffusion coefficient is 1/6 in lattice units. A laminar flow regime is solved in the whole domain with a fully developed flow condition enforced at the outlet. The concentration distribution is obtained by solving the advection-diffusion equation. Details about the solution of the model are available in Zhou et al. (2015). Figure 2b shows ion mass flux along the bottom wall of the channel obtained from both Zhou et al. (2015) and our present study. It is clear that both results are in excellent agreement.

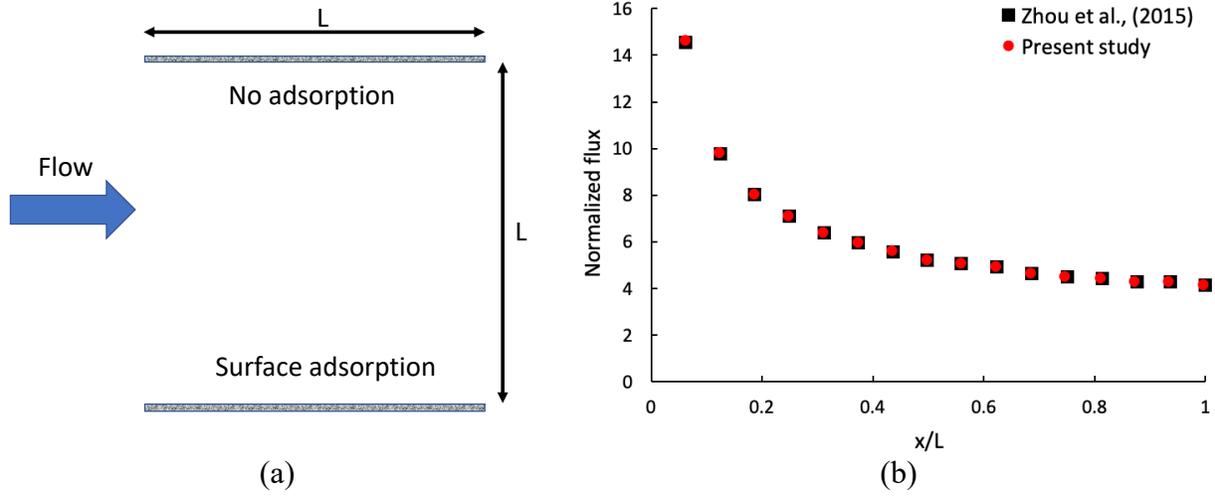

Figure 2 (a) The schematic of the channel with advection, diffusion and adsorption at the bottom; (b) comparison with the results from Zhou et al. (2015).

We validate our simulation of ion migration caused by electrical potential by comparing with the analytical solutions in a half channel shown in Figure 3(a). The width of the domain is 1 μm. Zero potential and constant concentration of $1\times10^{-2}$ mol/m$^3$ are applied at the centerline. The constant electrical potential of $-10$ mV is enforced at the bottom wall. The analytical solution of ion concentration is [40],

$$C_k = C_{k\infty}\exp{(-\frac{ez_k\psi}{K_BT})}, \quad (9)$$

where the ion valence $e$ is $1.6\times10^{-19}C$. The value of Boltzmann constant $K_B$ is $1.38\times10^{-23} JK^{-1}$. The analytical electrical potential distribution is computed as [22],

$$\psi(y) = 4\frac{K_BT}{e}tanh^{-1}(\tanh{(\frac{e\psi_o}{4K_BT}}\exp{(-\frac{y}{\sqrt{\frac{\varepsilon_r\varepsilon_0 K_BT}{2C_{k\infty}eF}}}))}). \quad (10)$$

The comparison of ion concentrations between our model and analytical solutions are demonstrated in Figure 3b. The simulation results of Na$^+$ and Cl$^-$ concentrations present good agreement with analytical solutions.

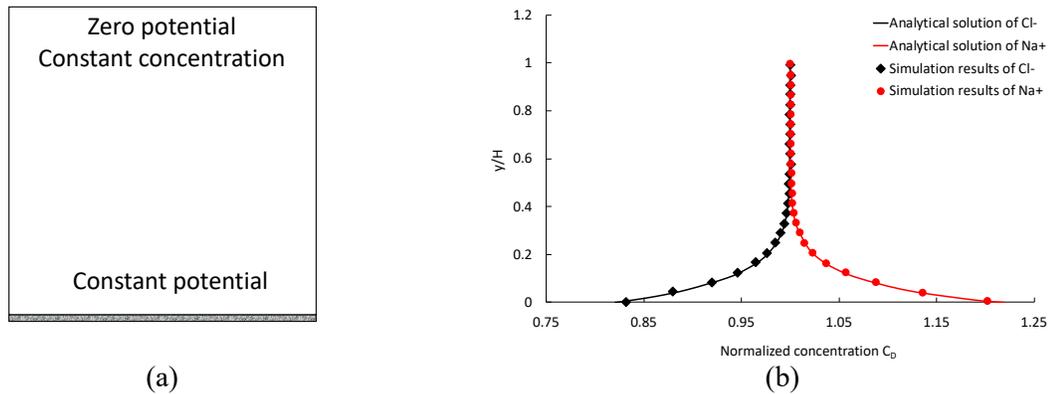

Figure 3 (a) Half channel geometry for validation of electro-migration;(b) Comparisons with analytical solutions for ion concentration

## 4. Ion transport and adsorption in porous electrodes at pore scale

In this section, pore-scale simulations of ion transport and adsorption are carried out in porous carbon-based electrodes. In the models, only the charging process of CDI is considered. To represent the real structures of the porous electrodes, a SEM image is digitalized for simulations (Figure 4). Ion transport and adsorption are simulated in the geometry of porous electrodes shown in Figure 4b. The digitized image for modeling has a size of 43.3 × 32.5 µm. The smallest pore has at least four lattice nodes, meeting the requirement of the LBM for accurate flow simulation [41]. Initially the system is saturated with $Na^+$ and $Cl^-$ and constant concentration ($C_0$=1.0) is enforced at the inlet (left). Fluid flows from left to right, and the ions are assumed to be adsorbed only on the surface of the solid. The diffusion coefficient of $Na^+$ and $Cl^-$ is set as $1.0 \times 10^{-9}\ m^2/s$ [42]. Under the effect of electrostatic force, the effective diffusion of $Cl^-$ is slower than the given one while the diffusion rate for $Na^+$ is higher. The adsorption constant is $K_a = 1.0 \times 10^{-6}\ m/s$ [22].

The dimensionless Peclet (Pe) and Damkohler (Da) numbers are used to describe the flow and adsorption conditions. The Peclet number, defined as $Pe = \frac{U l_c}{D}$, characterizes the strength of advection compared with molecular diffusion. The Damkohler (Da) number, defined as $Da = \frac{K_a l_c}{D}$, describes the ratio of adsorption rate constant over the molecular diffusion. The average velocity $U$ is calculated as $U = \frac{\sum_i u_i}{N \phi}$, where $N$ is the total number of grid blocks (image voxels) and $\phi$ is porosity. The characteristic length $l_c$ is calculated as $\pi/s$ [36], where $s$ is the specific surface. Because both molecular diffusivity and characteristic length are fixed in our simulations, the Peclet number is then used to describe the effect of flow rate.

### 4.1. The effect of Pe (flow velocity)

In this Section, the effect of Pe is quantitatively analyzed and investigated by carrying out simulations with different flow velocities but fixed adsorption coefficient (such that Da= $4.28 \times 10^{-3}$) and electrical potential (-40 mV).

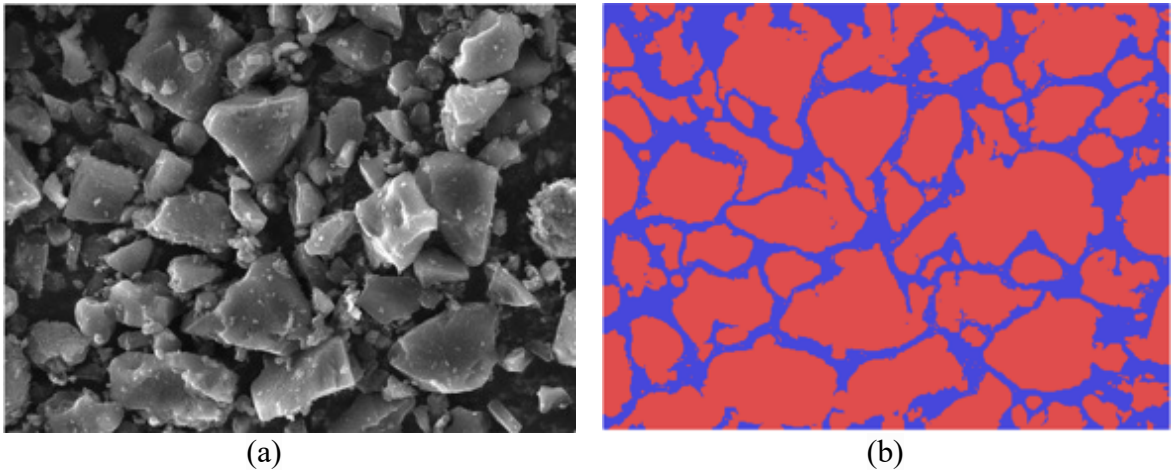

(a)  (b)

Figure 4 (a) SEM images (43.3 × 32.5 µm) of mesoporous carbon reproduced from Li et al. (2019) and (b) digitalized microstructure from SEM for modeling.

The steady-state concentration distributions of Na$^+$ at different flow velocities are presented in Figure 5. A gradient in Na$^+$ concentration (from inlet to outlet) is observed for all the cases as a result of combined transport and adsorption. However, there are significant variations in the Na$^+$ ion concentration distributions in porous electrodes at different Pe numbers. At low velocity, for example, in Figure 5a, most Na$^+$ is observed in fast flow paths. The concentration of Na$^+$ ions in small pore spaces (or narrow channels) are lower. In narrow pore space, the local velocity is lower while the impact of electrical attraction becomes stronger. This leads to more ions attracted to the solid surface and then absorbed. With the increase of velocity, at Pe= $1.03 \times 10^{-2}$ ($2.4 \times 10^{-6} m/s$) (Figure 5b), the concentrations of Na$^+$ are higher, particularly in the pores close to the outlet. This is because of the faster mass transfer resulting from stronger advection at higher Pe (velocity). For a fixed Da (adsorption rate constant), a faster mass transfer leads to more ions transported through the pore space in electrodes without being adsorbed. At the highest velocity in Figure 5c, Na$^+$ ions are transported to more narrow pore throats due to the faster mass transfer, where higher concentrations are observed.

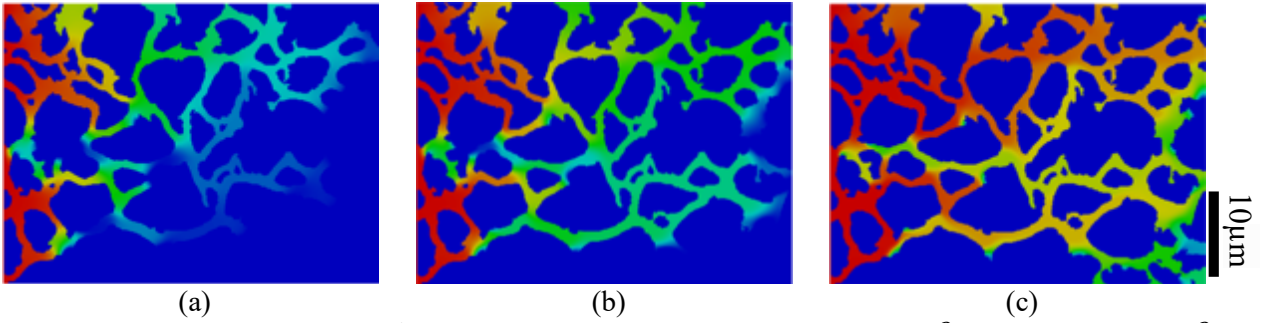

(a)      (b)      (c)

Figure 5 The concentration of Na$^+$ in the porous electrodes with Pe=$2.58 \times 10^{-3}$ (a); Pe=$1.03 \times 10^{-2}$ (b); Pe=$4.12 \times 10^{-2}$ (c).

The vertically averaged concentration of Na$^+$ ions as a function of distance from the inlet is shown in Figure 6. As the pore space is initially assumed to be saturated with salt water, the Na$^+$ concentration at the inlet is 1.0. With the increase of distance from the inlet, the concentration decreases due to mass transfer and adsorption for all the three cases. The Na$^+$ concentration is always higher at a higher Pe. This is because at a higher Pe (flow velocity), the mass transfer is faster so that more ions are flushed out the domain during the same period of time before getting adsorbed on the electrode surface. Higher concentration means higher water salinity and lower desalination efficiency.

We also compare the salt removal percentages of the porous electrodes at different Pe with same potential difference 40mV (Figure 7a). As expected, at higher Pe, lower percentage of initial salt ions are removed. This is because the adsorption is relatively slow compared to the advection so that more ions are transported to the outlet before getting adsorbed. Nearly 99% of the salt ions are removed at the outlet for the lowest Pe=$2.58 \times 10^{-3}$, as opposed to 85% for the highest Pe ($4.12 \times 10^{-2}$).

The total amount of salt ions adsorbed (or removed from the solution) within a certain period of time is calculated by $\frac{UAP_{rm}\Delta t}{V_t \phi}$, where $P_{rm} = \frac{C_{in}-C_{out}}{C_{in}}$ is the salt removal percentage, $A$ is the cross-section area and $V_t$ is the total volume of porous electrodes. Figure 7b shows the total amount of adsorbed ions as a function of Pe. For the purpose of comparison, we normalize the amounts of adsorbed ions by dividing the results at Pe=$1.03 \times 10^{-2}$ and potential diffetence 40mV. Due to the faster mass transfer at higher Pe, more ions are adsorbed in the same time period. The normalized amount is 3.7 at the highest Pe ($4.12 \times 10^{-2}$) while it is only 0.25 at Pe=$2.58 \times 10^{-3}$. This implies the effective rate of ion adsorption is greatly affected by the Pe number. Though the salt removal percentage is high at low Pe ($2.58 \times 10^{-3}$), fewer ions are adsorbed within the same time period because of the slow mass transfer.

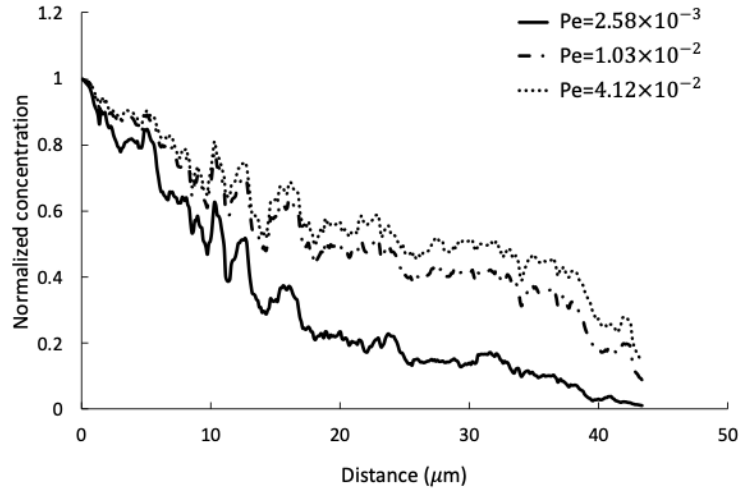

Figure 6 The vertically averaged concentration of Na$^+$ ions as a function of distance from the inlet

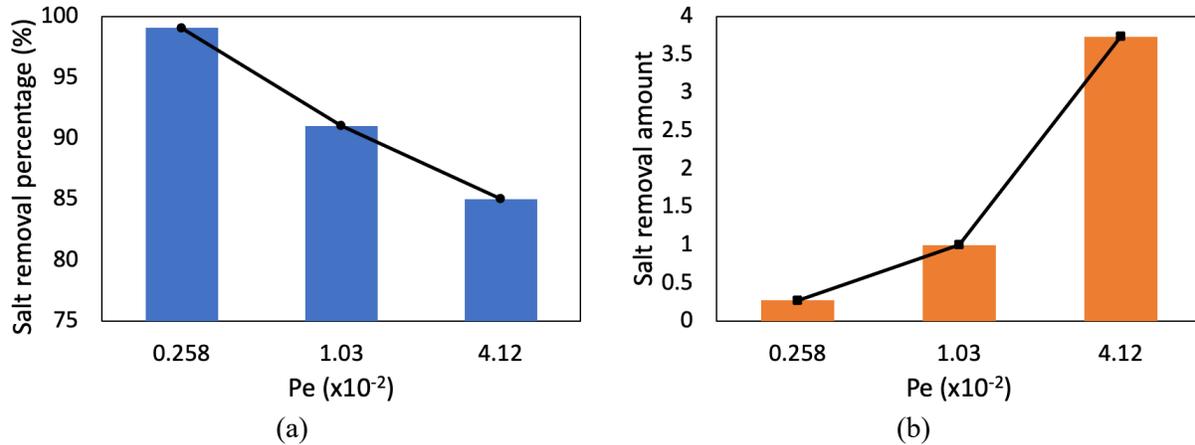

Figure 7 Salt removal percentage at different Pe with potential difference of 40mV (a) and total salt ions removed within the same period of time as a function of Pe (b)

## 4.2. The impact of electrical potential

Electrical potential is also an important parameter for CDI operation. An optimized potential helps save energy and improve desalination efficiency [9,21]. The impact of electrical potential on the desalination process is investigated in this section. Simulations are performed on the same geometry of the porous electrodes in section 4.1. Identical Pe ($1.03 \times 10^{-2}$) with three different electrical potential differences is applied in all the three cases. The steady-state Na$^+$ concentration distributions are shown in Figure 8. At the lowest potential difference of 20mV (Figure 8a), the electro-migration of ions is the slowest. Under such condition, more Na$^+$ ions are transported along the flow direction without being attracted to the solid surface and getting adsorbed. This leads to higher ion concentration in the pore space. When the potential difference is increased to $40mV$ (Figure 8b), Na$^+$ concentration in many narrow pore throats becomes much lower and even approaches zero. This indicates higher electrical potential difference accelerates the adsorption by causing stronger electrical attraction. Thus, more ions are transported to the solid surface leading to more adsorption. At the highest potential difference, as expected, the concentration of Na$^+$ is the lowest among these three cases due to the strong adsorption of salt ions, as can be seen in Figure 8 (c).

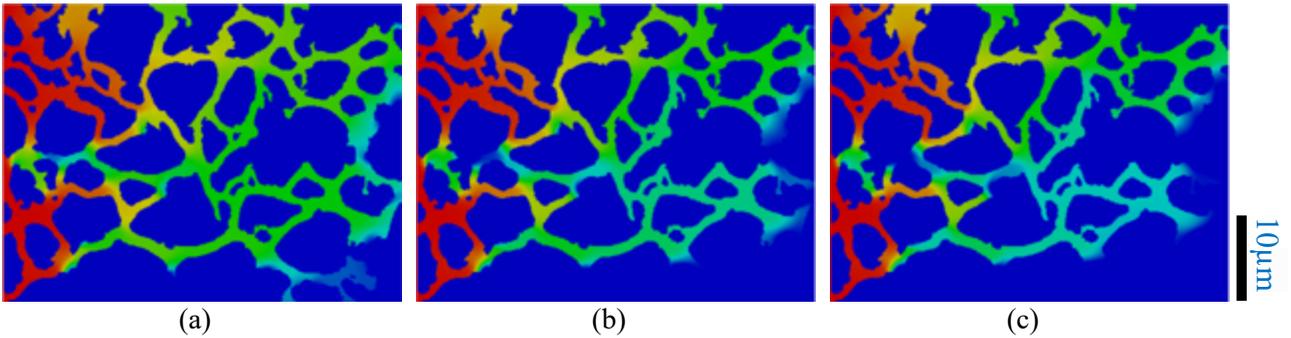

Figure 8 Na$^+$ concentration distribution with wall electrical potential (a) $-20mV$ (b) $-40m$V (c) $-60mV$ at Pe=$1.03 \times 10^{-2}$

The vertically averaged concentration of Na$^+$ ions as a function of distance from the inlet at different electrical potential differences are compared in Figure 9. As can be seen, the Na$^+$ concentration decreases with the distance, and it is always higher at the lower potential difference. At a high potential difference ($60mV$), the outlet Na$^+$ concentration is reduced to a quite low value (0.04). At the medium potential difference ($40mV$), the outlet ion concentration is 0.1. The highest outlet concentration is 0.15 when the potential difference is $20mV$. This is consistent with the findings in Figure 8 that larger potential difference can effectively enhance the ion adsorption and reduce the Na$^+$ ion concentration in electrodes. Higher potential difference is desirable operating conditions if high desalination efficiency is a priority. This result also shows that our model can be used to quantitatively describe the Na$^+$ concentration variations during adsorption at different electrical potential differences.

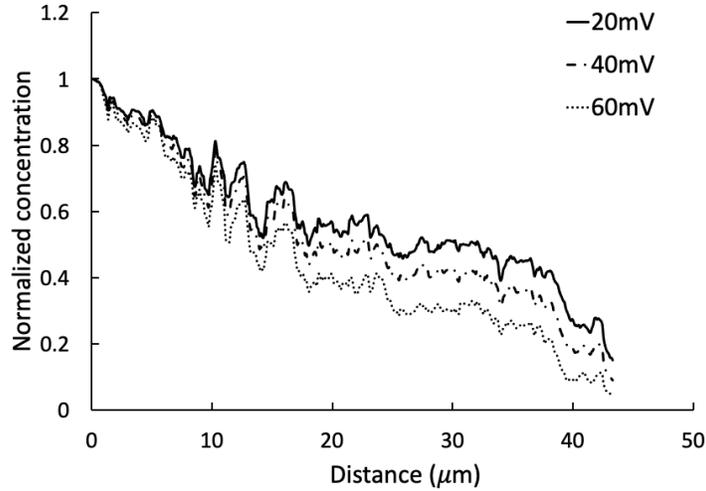

Figure 9 The vertically averaged concentration of Na$^+$ ions as a function of distance from the inlet at different potential differences for Pe=$1.03 \times 10^{-2}$

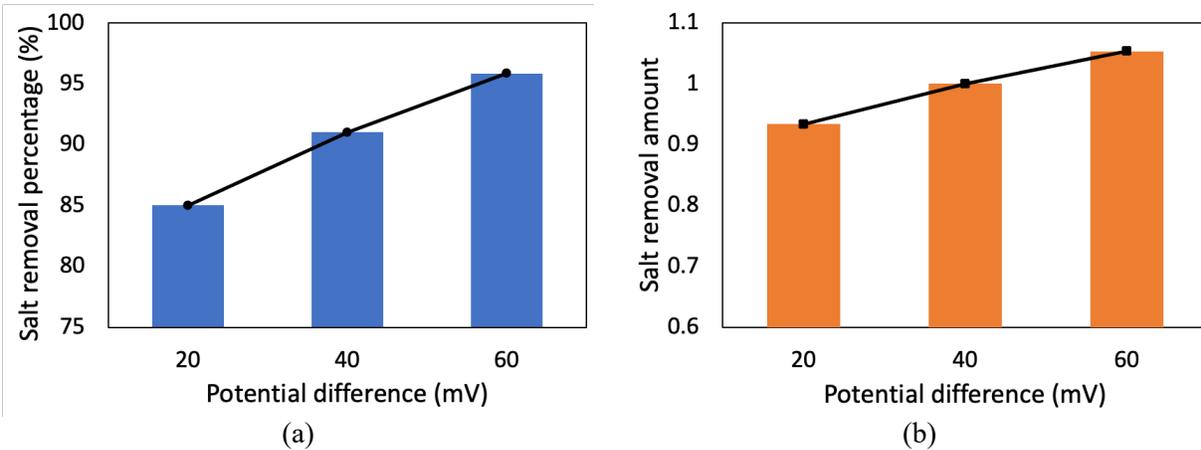

Figure 10 Salt removal percentage at different potential differences (a) and total salt ions removed within the same period of time as a function of potential difference (b)

The salt removal percentages are also compared at fixed Pe=$1.03 \times 10^{-2}$ but different electrical potential (Figure 10a). The salt ions removal percentage can reach 95% when potential difference is increased to 60mV. It is concluded that high potential difference results in high salt removal percentages.

As can be seen in Figure 10b, the amount of removed ions increases with the potential difference for a fixed Pe. This is because the potential difference enhances the ion migration to the solid surface for adsorption when the strength of advection remains the same. However, the difference is not significant among these three cases. When the potential difference is increased from 20 mV to 60 mV, the normalized salt removal amount only increases by 0.1. In comparison with the results at different Pe in Figure 7b, this suggests the salt removal rate is more sensitive to Pe numbers than to potential difference.

These results in Section 4.1 and 4.2 suggest that when determining the optimal CDI operating conditions, both salt removal percentage and ion removal rate should be considered to achieve a balance among desalination efficiency, water throughput, and cost. A comprehensive parametric study based on numerical simulations can facilitate the process.

## 5. 3D modelling of ion adsorption in fibrous electrodes

Fibers are popular adsorbent used in capacitive deionization electrodes [5,13,15]. Figure 11 shows a SEM image of polyacrylonitrile (PAN) fibers in a porous electrode. In this section, we create fibrous structures and perform simulations of ion transport and adsorption in 3D porous electrodes.

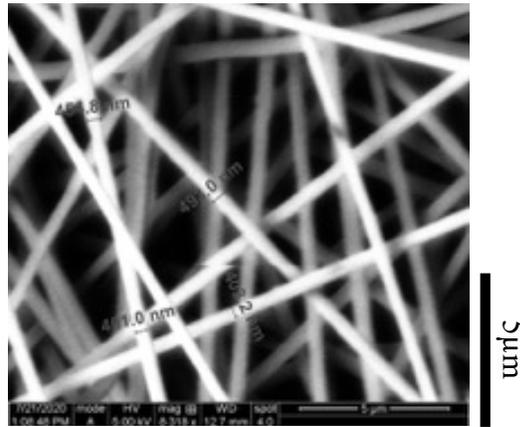

Figure 11 SEM image of PAN fibers in porous electrodes

To generate the fibrous media, we randomly place cylindrical fibers with given diameters in a 3D cubic. We first generate a vector containing spatial orientation of the fiber's core and the fiber diameter. Then, we extend the fiber core line from the origin point along the randomly determined orientation in both directions until it intersects the domain boundaries. The grid blocks or voxels that are within the radius of fiber are assigned as solid [43–45]. The radius of the fibers remains constant along the fiber core line and fibers are placed with free overlapping. The size of the domain is 150×150×150 voxels while the resolution of each voxel is 100nm. Figure 12 shows four samples of the reconstructed fibrous medium with the fiber diameter of 400nm and porosity from 0.5 to 0.8. The black color denotes the carbon fibers and void area the pore space.

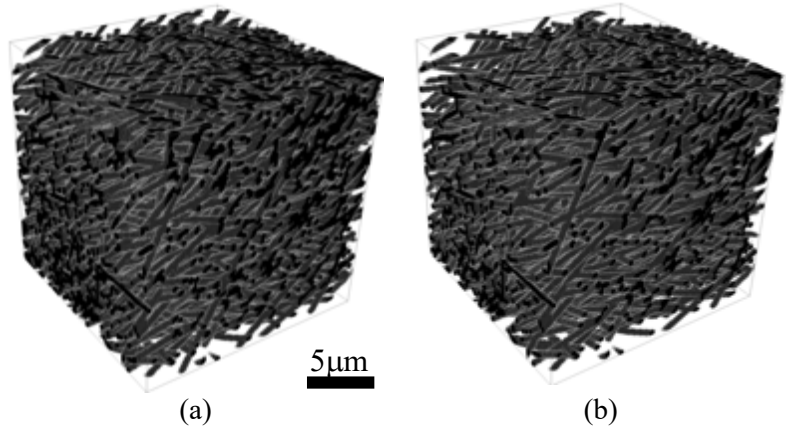

(a)          (b)

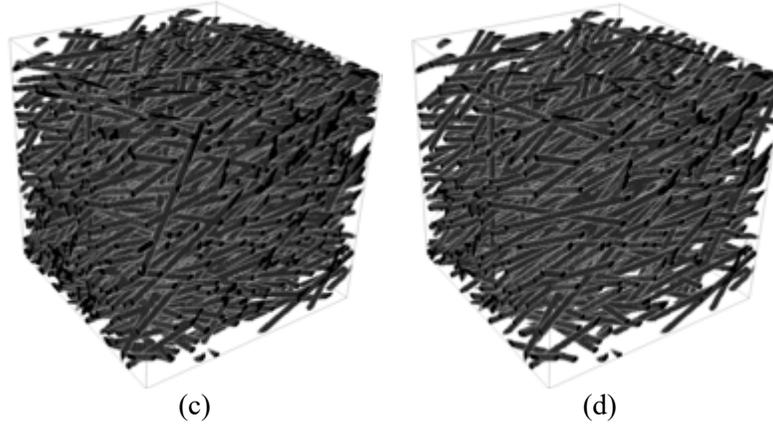

|(c)|(d)|

Figure 12 3D fibrous media with fiber diameter 400nm and porosity of (a) 0.5; (b)0.6; (c) 0.7 and (d) 0.8

We investigate the effect of porosity and fiber diameters on the ion adsorption behaviors in fibrous media. The normal potential difference used in CDI ranges from 40mV to 1.4V, which depends on the operating condition [9,12,22]. As the porosity and pore space is larger in these 3D fibrous media, we choose a higher potential difference 1.2V. The potential at the material surface is set as -1.2V and the average flow velocity is $1.0\times10^{-6}$ ms$^{-1}$. Figures 13-15 show the Na$^+$ concentration distributions in fibrous media with fiber diameter 400 nm, 600nm and 800nm, respectively. For each diameter, four porosity values (0.5, 0.6, 0.7 and 0.8) are considered. Figure 16 shows quantified salt removal percentage at different fiber diameters and porosities. For the same fiber diameter, higher porosity results in lower salt removal percentage. For example, in the simulations with fiber diameter of 600nm (black squares), the salt removal percentage decreases from 84% to 28% when the porosity increases from 0.5 to 0.8. The reason is that for a given fiber diameter, a lower porosity means denser fibers and higher specific surface area, which leads to stronger adsorption for a fixed flow rate.

For the simulations with same porosity but different fiber diameters, the results show salt removal percentage increases as the fiber diameter decreases. For example, at porosity 0.7, the salt removal percentage is calculated as 40% in fibrous media with a diameter of 800nm, but it can be as high as 62% when the fiber diameter is 400nm. Again, the reason is that for a given porosity, a smaller fiber diameter means higher specific surface, causing stronger adsorption in the porous electrodes when the same flow velocity is enforced. It is indicated the finer fibers can greatly improve the salt removal percentage.

Our results suggest that both porosity and fiber diameter are important parameters controlling desalination efficiency. A combination of lower porosity and finer fiber leads to higher desalination efficiency. However, this requires a much higher pressure gradient to keep up the flow rate, again requiring a trade-off among desalination efficiency, water throughput, and operation cost.

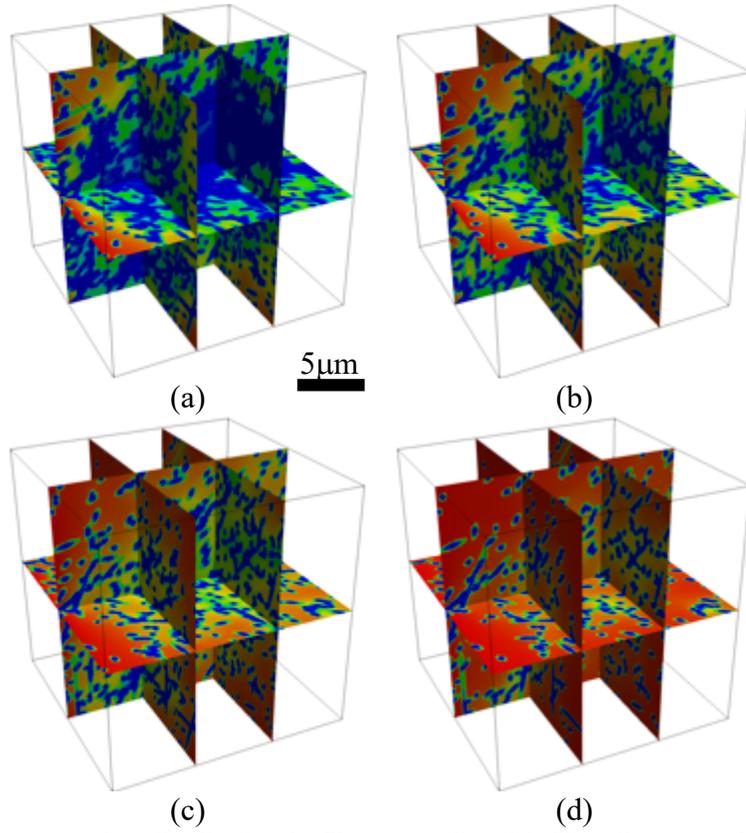

Figure 13 Na$^+$ concentration distribution in fibrous media with fiber diameter of 400nm at porosity (a) 0.5; (b) 0.6; (c) 0.7 and (d) 0.8

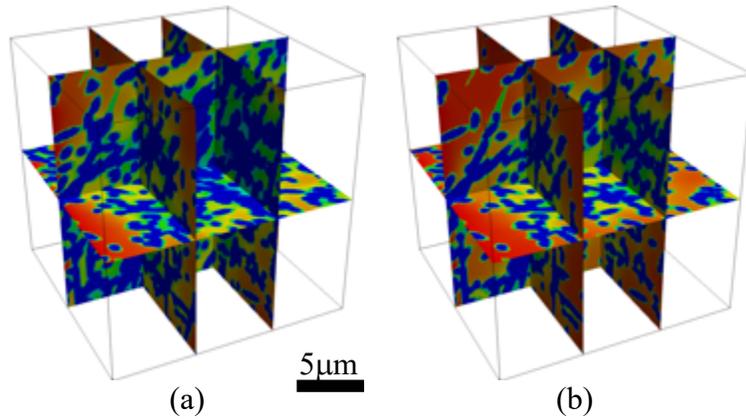

(a) (b)

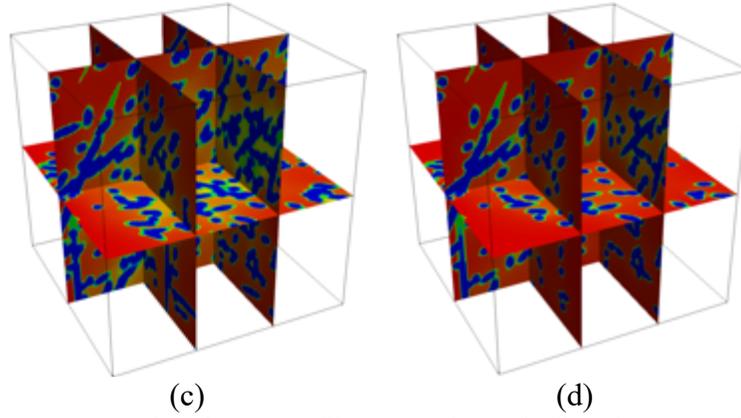
(c) (d)

Figure 14 Na$^+$ concentration distribution in fibrous media with fiber diameter of 600nm at porosity (a) 0.5; (b) 0.6; (c) 0.7 and (d) 0.8

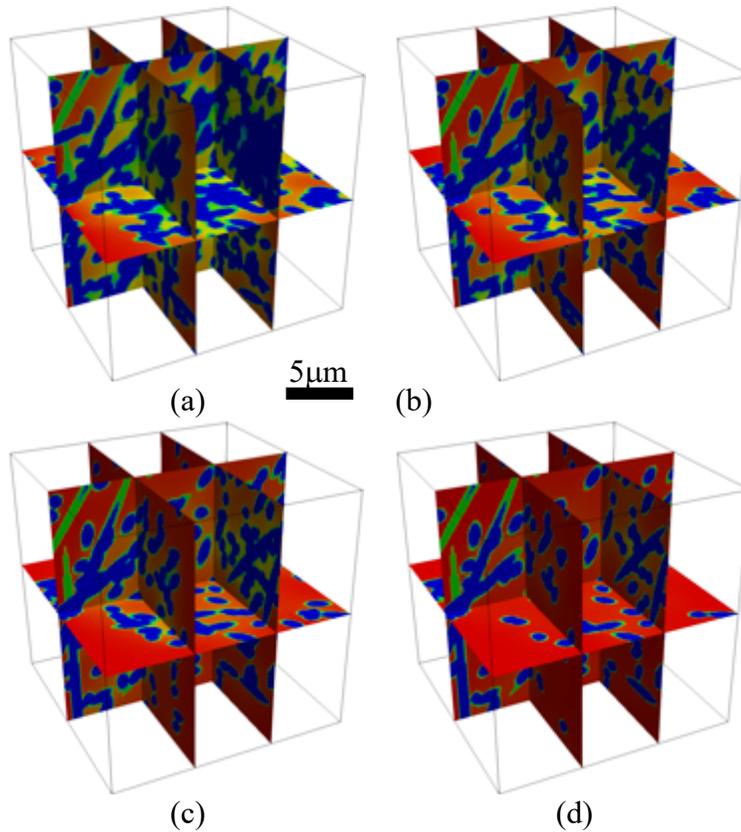
(a) 5μm (b)

(c) (d)

Figure 15 Na$^+$ concentration distribution in fibrous media with fiber diameter of 800nm at porosity (a) 0.5; (b) 0.6; (c) 0.7 and (d) 0.8

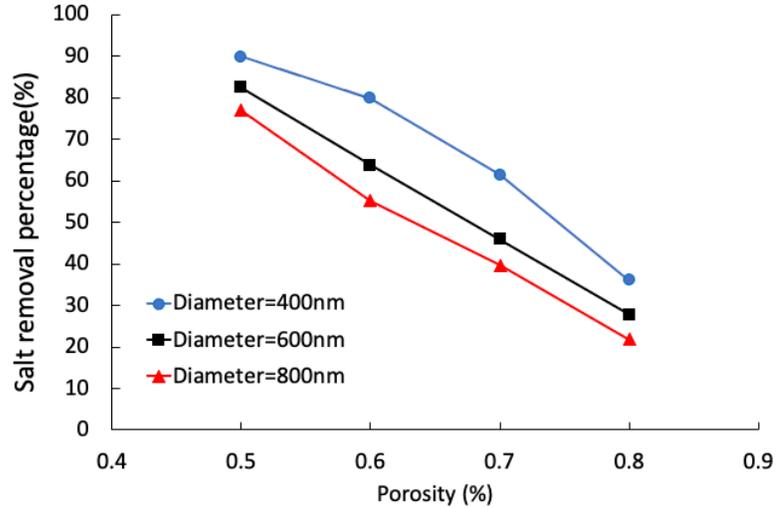

Figure 16 Salt removal percentage in fibrous media at different fiber diameters and porosities.

## 6. Conclusions

In this paper, we presented a pore-scale numerical study of the electrochemical process involving ion transport and adsorption in porous electrodes. Advection, diffusion, electrical migration and adsorption are all considered in the simulations. The model is validated by comparing results with published results and analytical solutions. The ion transport and adsorption demonstrate great dependency on flow velocity and electrical potential. The impact of velocity was studied by performing simulations on images of real porous electrodes. Our results show that higher velocity results in lower desalination efficiency when adsorption rate constant and potential difference are fixed. Simulations of varying potential differences were also conducted. The results indicate that larger potential difference can effectively enhance the ion adsorption and reduce the $Na^+$ ion concentration in electrodes. It also shows our model can be used to quantitatively describe the $Na^+$ concentration variations during adsorption at different electrical potential differences. This allows predicting optimal electrode parameters under different operating conditions.

We also performed simulations in 3D fibrous media. For the simulations with the same fiber diameter, the higher the porosity, the lower the obtained salt removal percentage, which indicates higher porosity results in weaker adsorption in porous electrodes. For the simulations with the same porosity but different fiber diameters, the salt removal percentage in fibrous media with finer fiber is higher than that with larger fiber diameters. It is indicated that the finer fibers enhance the adsorption in porous electrodes. We also found that a combination of lower porosity and finer fiber leads to higher desalination efficiency. However, this requires a much higher pressure gradient to maintain the desired flow rate.

The current work provides a numerical approach to better understand and quantitatively characterize ion transport and adsorption in porous electrodes, which can help the design and optimization of CDI electrodes and operating conditions for a trade-off among desalination efficiency, water throughput and cost.

## Nomenclature

| | | | |
|---|---|---|---|
| $\boldsymbol{u}$ | velocity vector | $\underline{n}$ | unit normal vector |
| p | pressure | $f$ | particle distribution function |
| μ | dynamic viscosity of the fluid | $e$ | microscopic velocity |
| $C$ | local concentration | $\Omega$ | collision operator |
| $D$ | diffusion coefficient | $f^{eq}$ | equilibrium function |
| $t$ | time | $\tau$ | relaxation time |
| $\upsilon$ | kinetic viscosity | $\upsilon$ | kinematic viscosity |
| $z$ | ion valence | $Pe$ | Peclet number |
| $E$ | elementary charge | $Da$ | Damkohler number |
| $K_B$ | Boltzmann constant | $U$ | average velocity |
| $\psi$ | electrical potential | $l_c$ | characteristic length |
| $N$ | total number of grid blocks | $s$ | specific surface |
| $T$ | temperature | $\phi$ | porosity |
| $\rho_e$ | net charge density | $P_{rm}$ | salt removal percentage |
| $\varepsilon_0$ | vacuum electric permittivity | $V_t$ | total volume of domain |
| $\varepsilon_r$ | dielectric constant | $A$ | cross-section area |
| $K_a$ | the adsorption constant | | |

## Acknowledgments

Research presented in this article was supported by the Laboratory Directed Research and Development (LDRD-ER) program of Los Alamos National Laboratory (LANL).